# Magnetism in Mn delta-doped cubic GaN: density-functional theory studies

Xingtao Jia[1], Wei Yang[2], Minghui Qin[3], Xinglai Zhang[4], Xindong Cui[4] and Mingai Sun[4]

[1]*College of Chemistry and Chemical Engineering, China University of Petroleum, Dongying, 257061, China*

[2]*College of Information, Shanghai Ocean University, Shanghai, 200090, China*

[3]*National Laboratory of Solid State Microstructures and Department of Physics, Nanjing University, Nanjing 210093, China*

[4]*College of Physics Science and Technology, China University of Petroleum, Dongying, 257061, China*

**Abstract** - The magnetism in 12.5% and 25% Mn delta-doped cubic GaN has been investigated using the density-functional theory calculations. The results show that the single-layer delta-doping and half-delta-doping structures show robust ground state half-metallic ferromagnetism (HMF), and the double-layer delta-doping structure shows robust ground state antiferromagnetism (AFM) with large spin-flip energy of 479.0 meV per Mn-Mn pair. The delta-doping structures show enhanced two-dimensional magnetism. We discuss the origin of the HMF using a simple crystal field model. Finally, we discuss the antiferromagnet/ferromagnet heterostructure based on Mn doped GaN.

Nitride-based magnetic semiconductors have attracted much attention recently for the potential above room temperature spintronics applications [1-13]. Among them, GaN based magnetic semiconductors are more attractive for the highest Curie temperature ($T_C$) ferromagnetism (FM) [1]. However, although many researches have reported successful synthesis of the nitride-based magnetic semiconductors, the results are hard to reproduce and always under controversy. To achieve applicable magnetism, magnetic impurities should be highly ordered patterned in the magnetic semiconductors. Unfortunately, the intrinsic aggregation trend of the magnetic impurities would deteriorate and even destroy the pattern map, and finally destroy the magnetism [14-19]. One-way to controllable pattern the magnetic impurities into the matrix semiconductor would be delta-doping [20-22], which can lead to metastable two-dimensional magnetism. Moreover, the delta-doping structures always show enhanced magnetism. Compared with homogeneous doping, the delta-doping is more practical via the molecular beam epitaxy (MBE) synthesis. Moreover, it is easy to achieve accurate single-, multi-, and fractional-layer delta-doping structures by modulating the flux of the precursors and time in the metastable MBE synthesis.

Here, we investigate the magnetism in the 25% and 12.5% Mn delta-doped cubic GaN using density-functional theory calculations. There are two stable crystal structures for GaN. Under ambient temperature, the wurtzite (hexagonal) phase is more stable than the zinc blende (cubic) phase. And, the cubic phase can be stabilized via doping with high concentration 3d-transition metal ions [23] or epitaxial growth on the appropriate matrix.

[24-27] Moreover, the study shows that the higher symmetrical crystal structures always show enhanced magnetism than the lower symmetrical ones. [28] In the studies, we find that the single-layer delta-doping and half-delta-doping structures show robust ground state half-metallic ferromagnetism (HMF), and the double-layer delta-doping structure shows robust grond state antiferromagnetism (AFM). Finally, we speculate the application of the antiferromagnet/ferromagnet bilayers based on Mn doped GaN.

We perform spin-polarized total energy density-functional calculations in the Perdew-Burke-Ernzerhof (PBE) generalized gradient approximation [29] using the plane-wave ultrasoft pseudopotential method [30] as implemented in the Cambridge Serial Total Energy Package (CASTEP) code [31]. For the 12.5% Mn single-layer delta-doped and 25% Mn double-layer delta-doped cubic GaN, a cell containing 16-fold single atom layers along [001] direction with lattice parameter a=b=3.086 Å and c=17.456 Å is used. For the 12.5% Mn half-delta-doped cubic GaN, a cell containing 8-fold 4 atoms layers with lattice parameter a=b=6.172 Å and c=8.728 Å is used. An energy cutoff of 450 eV is used in all calculations. Brillouin zone integrations are preformed with the special k-point method over a 8x8x2 Monkhorst-Pack mesh for the single-layer delta-doping and double-layer delta-doping structures and a 4x4x3 mesh for the single-layer half-delta-doping structure. Both the lattice parameters and atomic positions are fully relaxed in all calculations. Through the letter, the formation energy of Mn substitution ($Mn_{Ga}$) in GaN is calculated with the well-established formula [32], which is defined as

$$E_f[Mn_{Ga}] = E[Mn_{Ga}] - E[bulk] - n\mu_{Mn} + n\mu_{Ga} \qquad (1)$$

where $E[Mn_{Ga}]$ and $E[bulk]$ is the total energy of the Mn doped GaN and pure GaN cell respectively. $\mu_{Mn}$ and $\mu_{Ga}$ is the chemical potentials of antiferromagnetic α-Mn and orthorhombic gallium (Ga) respectively.

The calculated magnetic ground state, ferromagnetic stabilization energy (energy difference between antiferromagnetic and ferromagnetic states, $\Delta E_{AF} = E_{AFM} - E_{FM}$), and $Mn_{Ga}$ formation energy in the 12.5% Mn single-layer delta-doped and half-delta-doped cubic GaN, and 25% Mn single-layer delta-doped and double-layer delta-doped cubic GaN are given in table. 1. Apparently, the single-layer delta-doping and half-delta-doping structures show robust ground state HMF, and the double-layer delta-doping structure shows robust ground state AFM. We study two types of the half-delta-doping structures. Therein, the homogeneous one shows $Mn_{Ga}$ formation energy of 1.758 eV, which is notably large than the heterogeneous one (1.534 eV). For the single-layer delta-doping structures, the lower doping concentration one (12.5%) show large $\Delta E_{AF}$ and lower $Mn_{Ga}$ formation energy than the higher doping concentration one (25%). The double-layer delta-doping structure shows the lowest $Mn_{Ga}$ formation energy among all studied structures, and robust ground state AFM with spin-flip energy of 479.0 meV per Mn-Mn pair (here, the paraferromagnetic state is about 320.0 meV higher than the ferromagnetic state). Thus for, we observe robust HMF and AFM in the Mn delta-doped cubic GaN.

In fig. 1, we give the spin-resolved density of states (DOS) of the 12.5% Mn homogenously doped and single-layer delta-doped GaN. Apparently, both structures show pronounced majority HMF around the Fermi level ($E_F$). It originates from the Hund rule that the majority $3d$ impurity states are full occupied. Generally, the DOS curves of the homogenous doping structure are curvous with many sharp peaks, while the peaks are broadened for the delta-doping structure. Around the $E_F$, the majority DOS of the delta-doping structure extends to a much higher energy than the homogenous doping one, while the latter shows larger DOS at the $E_F$. Besides, the DOS of the two structures in the band gap of the bulk GaN are also different. For the delta-doping structure, there are two clusters of impurity DOS peaks above the top of the valence band separated by the pseudogap, which can be associated with the $e_g$-$t_{2g}$ splitting expected from a simple crystal field model of the bonding in cubic GaN. But, the $e_g$ and $t_{2g}$ states are melting into one peaks cluster in the homogeneous doping structure. The differences are come from the existence of John-Teller distortion in the delta-doping structure. Obviously, both Mn and N contribute to the majority HMF around the $E_F$, evidencing strong hybridization between Mn and N atoms.

As well as a high $T_C$, a half-metallic electronic structure and good compatibility with the mainstream semiconductors also are the key factors to the ideal spintronics materials. However, whether the HMF can be maintained up to the $T_C$ or not is determined by the electronic structure. According to Hordequin et al. [33] and Qian et al., [34] the width of the spin-flip gap $\delta$ is proportional to the electronic phase transition temperature $T^*$ from half-metal to normal ferromagnet. Here, for the delta-doping structure in fig. 1, the spin-flip gap $\delta$ is defined by the distance between the bottom of the conduction band in the minority DOS and the Fermi energy; however, for the homogeneous doping structure, which is defined by the distance between the uppermost majority DOS and the Fermi energy. We can see that the homogeneous doping structure shows a larger minority band gap but a smaller spin-flip gap; correspondingly, the delta-doping structure shows a smaller minority band gap but a larger spin-flip gap. So, we can deduce that the latter would own higher $T^*$ than the former. Furthermore, the DOS at higher energy (above the $E_F$) is important to the hot-electron technology. For the homogeneous doping structure, the DOS show electronic phase of majority HMF, semiconductor, minority HMF, and minority FM in turn, when elevates the $E_F$ to higher energy. However, for the delta-doping structure, the sequence is majority HMF, minority FM, minority HMF, and minority FM.

   The formation mechanism of the HMF in Mn doped cubic GaN can be dictated by a crystal field model. In fig.2, we give a schematic description of the *p-d* hybridization of Mn in N tetrahedron ($T_d$) field. Therein, the Mn-*d* electrons would experience an $e_g$-$t_{2g}$ splitting under the N $T_d$ field, which would hybridize with the neighbor N-*p* further and then result in the HMF. For symmetrical reason, the Mn-$t_{2g}$ orbitals would hybridize with N-*p* strongly while Mn-$e_g$ is rather weak. So, the Mn-$e_g$ orbitals would keep rather non-bonding. The half-metallic gap is determined by the energy difference between the minority bonding *p-d* hybrids orbitals and Mn-$e_g$. So, any factor strengthened the *p-d* hybridization would broaden the spin-flip gap $\delta$ and enhance the HMF.

   To study the magnetic interaction between the Mn impurities, we give the sites-resolved magnetic moments of the 12.5% Mn single-layer delta-doped GaN and 25% Mn double-layer delta-doped GaN in fig. 3. For the single-layer delta-doping structure, the magnetic impurity Mn, first-neighbor N, second-neighbor Ga, and third-neighbor N show magnetic moments of 4.00, -0.04, 0.02 and 0.02 $\mu_B$ respectively; and the other atoms show zero magnetic moments. Evidently, the system shows a two-dimensional FM character. When removing all atoms except Mn, the cell shows ground state AFM. So, we can see that the magnetic double exchange interaction would be responsible for the ground state two-dimensional HMF [19]. In the studies, we define two AFM states for the double-layer delta-doping structure, one is defined by alternating planes spin up/spin down in the direction [001] of cubic GaN, another is defined by spin up and down in the same plane perpendicular to [001]. The results show that the latter is more energetically favorable, which is different with the study about the MnN layers embedded in hexagonal GaN [35]. As shown in fig. 3, only the magnetic impurity Mn, first- and second-neighbor Ga show non-zero magnetic moments. This is different with the low magnetic moment at the Mn site in the bulk cubic MnN [36]. Obviously, the double-layer delta-doping structure shows two-dimensional AFM.

The antiferromagnet/ferromagnet heterostructure is the bases of many spintronics devices such as spin valve, giant magnetoresistance (GMR) magnetic sensors, and magnetic tunnel junction. As discussed above, we can achieve the antiferromagnet/ferromagnet bilayers just by modulating the number of the Mn layers in the GaN matrix. This is a valuable character, which would make the structure more concise and the synthesis easier.

In conclusion, we investigate the magnetism in the 12.5% Mn single-layer detla-doped and half-delta-doped cubic GaN, and 25% Mn double-layer delta-doped cubic GaN using the density-functional theory calculations. The results show that the single-layer delta-doping and half-delta-doping structures show robust ground state HMF, and the double-layer delta-doping structure shows robust ground state AFM with large spin-flip energy of 479.0 meV per Mn-Mn pair. Moreover, our studies show the delta-doping structures demonstrate enhanced two-dimensional HMF than the homogeneous doping structure. The origin of HMF is discussed using a crystal filed model, and the double-exchange magnetic interaction should be responsible to the ground state HMF in Mn doped GaN. Finally, we discuss the antiferromagnet/ferromagnet heterostructure based on Mn doped GaN.

***

The work is partially supported by the Postgraduate Innovation Foundations of China University of Petroleum (No. B2008-8).

# REFERENCES

[1] Sonoda S., Shimizu S., Sasaki T., Yamamoto Y. and Hori H., *J. Cryst. Growth*, **91** (2002) 7911.
[2] Sanyal B., Bengone O. and Mirbt S., *Phys. Rev. B*, **68** (2003) 205210.
[3] Thaler G. T., Overberg M. E., Gila B., Frazier R., Abernathy C. R., Pearton S. J., Lee J. S., Lee S. Y., Park Y. D., Khim Z. G., Kim J. and Ren F., *Appl. Phys. Lett.*, **80** (2002) 3964.
[4] Sasaki T., Sonoda S. and Yamamoto Y., *J. Appl. Phys.*, **91** (2002) 7911.
[5] Hori H., Sonoda S., Sasaki T., Yamamoto Y., Shimizu S., Suga K. -I. and Kindo K., *Physica B*, 324, (2002) 142.
[6] Liu C., Yun F. and Morkoc H., *J. Mater. Sci.: Mater. Electron.*, **16** (2005) 555.
[7] Liu H. X., Wu S. Y., Singh R. K., Gu L., Smith D. J., Newman N., Dilley N. R., Montes L. and Simmonds M. B., *Appl. Phys. Lett.*, **85** (2004) 4076.
[8] Yoshii S., Sonoda S., Yamamoto T., Kashiwagi T., Hagiwara M., Yamamoto Y., Akasaka Y., Kindo K. and Hori H., *Europhys. Lett.*, **78** (2007) 37006.
[9] Hynninen T., Raebiger H., Boehm J. von and Ayuela A., *Appl. Phys. Lett.*, **88** (2006) 122501.
[10] Pearton S. J., Abernathy C. R., Thaler G. T., Frazier R. M., Norton D. P., Ren F., Park Y. D., Zavada J. M., Buyanova I. A., Chen W. M. and Hebard A. F., *J. Phys.: Condens. Matter*, **16** (2004) R209.
[11] Bonanni A., *Semicond. Sci. Technol.*, **22**, (2007) R41.
[12] Bouzerar G., Ziman T. and Kudrnovsky J., *Europhys. Lett.*, **69** (2005) 812.
[13] Jungwirth T., Sinova J., Masek J., Kucera J., Macdonald A. H., *Rev. Mod. Phys.*, **78** (2006) 809.
[14] Cui X. Y., Medvedeva J. E., Delley B., Freeman A. J., Newman N. and Stampfl C., *Phys. Rev. Lett.*, **95** (2005) 256404.
[15] Miura Y., Shirai M. and Nagao K., *J. Phys.: Condens. Matter*, **16** (2004) S5735.
[16] Kaspar T. C., Droubay T., Heald S. M., Engelhard M. H., Nachimuthu P. and Chambers S. A., *Phys. Rev. B*, **77** (2008) 201303.
[17] Collins B. A., Chu Y. S., He L., Zhong Y. and Tsui F., *Phys. Rev. B*, **77** (2008) 193301.
[18] Giraud R., Kuroda S., Marcet S., Bellet-Amalric E., Biquard X., Barbara B., Fruchart D., Ferrand D., Cibert J. and Mariette H., *Europhys. Lett.*, **65** (2004) 553.
[19] Sandratskii L. M., Bruno P. and Mirbt S., *Phys. Rev. B*, **71** (2005) 045210.
[20] Qian M. C., Fong C. Y., Liu K., Pickett W. E., Pask J. E. and Yang L. H., *Phys. Rev. Lett.*, **96** (2006) 027211.
[21] Nazmul M., Amemiya T., Shuto Y., Sugahara S. and Tanaka M., *Phys. Rev. Lett.*, **95** (2005) 017201.
[22] Jeon H. C., Kang T. W., Kim T. W., Kang J. and Chang K. J., *Appl. Phys. Lett.*, **87** (2005) 092501.
[23] Choi E. -A. and Chang K.J., *Physica B*, **401-402** (2007) 319.
[24] Strite S., Ruan J., Li Z., Salvador A., Chen H., Smith D. J., Choyke W. J., Morkoc H., *J. Vac. Sci. Technol. B*, **9** (1991) 1924.
[25] Wang D., Hiroyama Y., Tamura M., and Ichikawa M., Yoshida S., *Appl. Phys. Lett.*, **76** (2000) 1683.
[26] Kurobe T., Sekiguchi Y., Suda J., Yoshimoto M. and Matsunami H., *Appl. Phys. Lett.*, **73** (1998)

2305.

[27] Daudin B., Feuillet G., Hubner J., Samson Y., Widmann F., Philippe A., Bru-Chevallier C., and Guillot G., Bustarret E., Bentoumi G. and Deneuville A., *J. Appl. Phys.*, **84** (1998) 2295.

[28] Jia X., Yang W., Li H. and Qin M., *J. Phys. D: Appl. Phys.*, **41** (2008) 115004.

[29] Perdew J. P., Burke K. and Ernzerhof M., *Phys. Rev. Lett.*, **77** (1996) 3865.

[30] Vanderbilt D., *Phys. Rev. B*, **41** (1990) 7892.

[31] Segall M. D., Lindan P. L. D., Probert M. J., Pickard C. J., Hasnip P. J., Payne S. M. C. and Clark J., *J. Phys.:Condens. Matter*, **14** (2002) 2717.

[32] Laks D. B., Van de Walle C. G., Neumark G. F., Blochl P. E. and Pantelides S. T., *Phys. Rev. B*, **45** (1992) 10965.

[33] Hordequin C., Ristoiu D., Ranno L., Pierre J., *Eur. Phys. J. B*, **16** (2000) 287.

[34] Qian M. C., Fong C. Y., Liu K., Pickett W. E., Pask J. E., Yang L. H., *Phys. Rev. Lett.*, **96**, (2006) 027211.

[35] Marques M., Scolfaro L. M. R., Teles L. K., Furthmüller J., Bechstedt F. and Ferreira L. G., *Appl. Phys. Lett.*, **88** (2006) 022507.

[36] Janotti A. and Wei S. H., and Bellaiche L., *Appl. Phys. Lett.*, **82** (2003) 766.

Table 1: Calculated magnetic ground state, ferromagnetic stabilization energy ($\Delta E_{AF}$), and $Mn_{Ga}$ formation energy ($E_f$) of 12.5% Mn full single-layer delta-doped and half-delta-doped cubic GaN, and 25% Mn double-layer delta-doped cubic GaN.

| | Ground state | $E_{AF}$ (meV) | $E_f$ (eV) |
|---|---|---|---|
| $Ga_{0.875}Mn_{0.125}N$[a] | HMF | 107.2 | 1.363 |
| $Ga_{0.875}Mn_{0.125}N$[b] | HMF | 153.4 | 1.534 |
| $Ga_{0.75}Mn_{0.25}N$[c] | AFM | -479.0 | 0.969 |
| $Ga_{0.75}Mn_{0.25}N$[a] | HMF | 92.0 | 1.369 |

[a] single-layer delta-doping
[b] single-layer half-delta-doping
[c] double-layer delta-doping

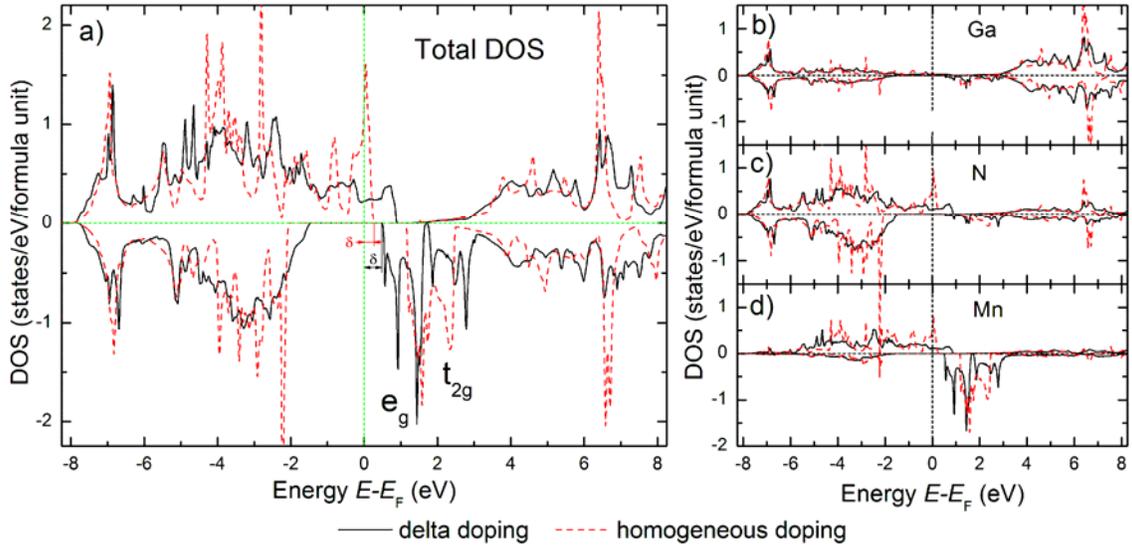

Fig. 1: (a) The spin-resolved density of states (DOS) of 12.5% Mn homogeneously doped and delta-doped cubic GaN. (b)-(d) The spin- and sites-resolved DOS for Ga, N, and Mn, respectively.

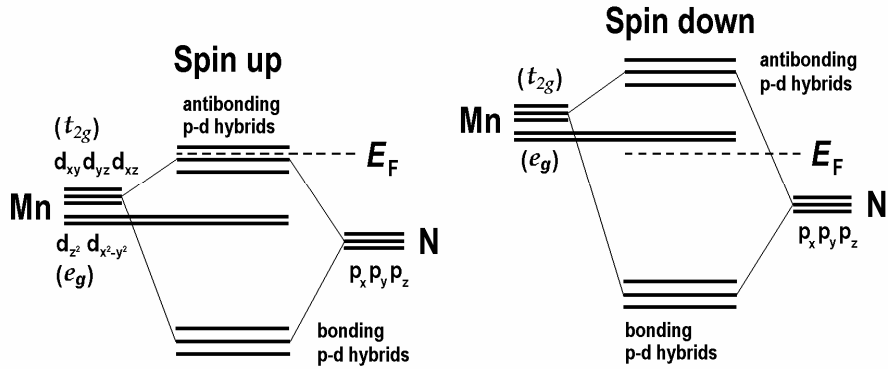

Fig. 2. The scheme of the *p-d* hybridization in Mn doped cubic GaN.

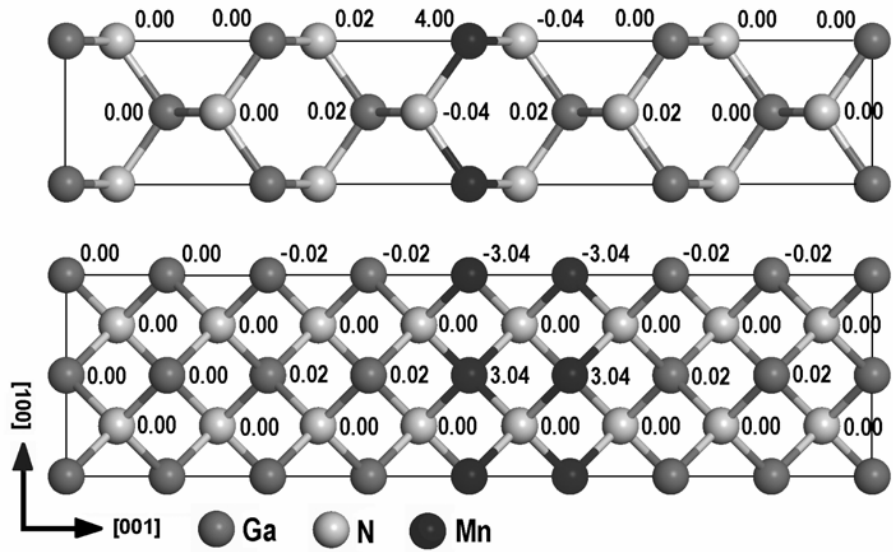

Fig. 3. The sites-resolved magnetic moments of the ferromagnetic 12.5% Mn single-layer (top) and antiferromagnetic 25% Mn double-layer delta-doped cubic GaN (bottom).